\documentclass[twocolumn,pra,aps,showpacs]{revtex4}
\usepackage{amsfonts}
\usepackage{amssymb}
\usepackage{epsfig}
\usepackage{amsbsy}
\usepackage{amssymb}
\usepackage{amsmath}
\usepackage{graphicx}
\usepackage{graphics}
\usepackage{xcolor}

\usepackage[colorlinks=true,linkcolor=blue,citecolor=blue]{hyperref}%,hidelinks,pdfborder={0 0 0},
\begin{document}

\title{Photon-aided and photon-inhibited Tunneling of Photons}
\author{ Xuele Liu}
\altaffiliation{xuele@okstate.edu}
\author{G.S. Agarwal}
\affiliation{Department of Physics, Oklahoma State University, Stillwater, Oklahoma
74078, USA}
\date{\today }

\begin{abstract}
In light of the interest in the transport of single photons in arrays of
waveguides, fiber couplers, photonic crystals, etc., we consider the quantum
mechanical process of the tunneling of photons through evanescently or
otherwise coupled structures. We specifically examine the issue of tunneling
between two structures when one structure already contains few photons. We
demonstrate the possibility of both photon-aided and photon-inhibited
tunneling of photons. The bosonic nature of photons enhances the tunneling
probability. We also show how the multiphoton tunneling probability can be
either enhanced or inhibited due to the presence of photons. We find similar
results for higher-order tunneling. Finally, we show that the presence of a
squeezed field changes the nature of tunneling considerably.
\end{abstract}

\maketitle

\section{Introduction}

The quantum transport properties of fermions have been extensively studied
in the last century \cite{bk1, bk2}. One interesting phenomenon in quantum
transport of fermions is the Coulomb blockade effect \cite{bk3}, which
happens when electrons are transported through a quantum dot. The Coulomb
blockade effect, arising from the Pauli exclusion principle, inhibits
tunneling by the presence of electrons in the quantum dot. The differences
between energy levels in small quantum dots are very large (with the
unoccupied higher level far from the Fermi energy); therefore, the transport
probability is very small and the tunneling is blocked.

An analog of the Coulomb blockade \cite{r1,r2} was demonstrated by Birnbaum
\emph{et al.} \cite{r3}, showing that if an atom resonant with a strongly
coupled single mode cavity could absorb one photon, then the absorption of a
second photon was inhibited. A similar experiment was reported in the
context of superconducting qubits \cite{r4}. Another phenomenon is the
dipole blockade \cite{r5} in Rydberg atoms where the excitation of a second
atom to a Rydberg state is forbidden if one atom is already excited to a
Rydberg state. The first excitation makes the second excitation non-resonant
which leads to the blockade effect, and is similar to the Coulomb blockade
in that it is based on the energy gap.

The inhibited tunneling of photons that we discuss in this paper has a more
fundamental origin---the bosonic statistics of photons, just like the Fermi
statistics in the Coulomb blockade. To understand this, we examine the
photon tunneling between two coupled single modes, with each mode referring
to a different structure.
%The tunneling we discuss is like the tunneling between different potential wells and is a coherent process  (\ref{Fig0}(a)).
This can be realized by coupled single-mode waveguide devices [Fig. \ref%
{Fig0}(a)] \cite{wg1,wg2}, fiber couplers, coupled resonators, and
optomechanical systems\cite{wg3}. The waveguides coupled by evanescent
fields can be tailored by changing the distance between waveguides \cite%
{wg1,wg2}. The evanescent coupling is responsible for the tunneling of
photons. This is equivalent to tunneling between two potential wells [Fig. %
\ref{Fig0}(b)]; the tunneling that we discuss is a coherent process.

In this paper we report photon-aided tunneling (PAT) and photon-inhibited
tunneling (PIT), which can occur due to the presence of photons in the other
waveguide. We find that even when the energy gap between the two waveguides
is large and the single-photon tunneling is negligible, the PAT is
significant and becomes about $1/\mathrm{e}$ when the number of photons in
the other waveguide is large. PIT occurs when the energy gap is small. When
the energy gap is zero, the tunneling rate without any photons in the other
waveguide is $100\%$; tunneling is totally inhibited when photons are
present in the other waveguide. Both PAT and PIT depend on the bosonic
nature of the photons. They are not allowed for electrons due to the Fermi
statistics. Our results are exact and go far beyond the perturbation theory.

\section{Model and method}

\label{s2}

\begin{figure}[t]
\includegraphics[width=0.28\textwidth]{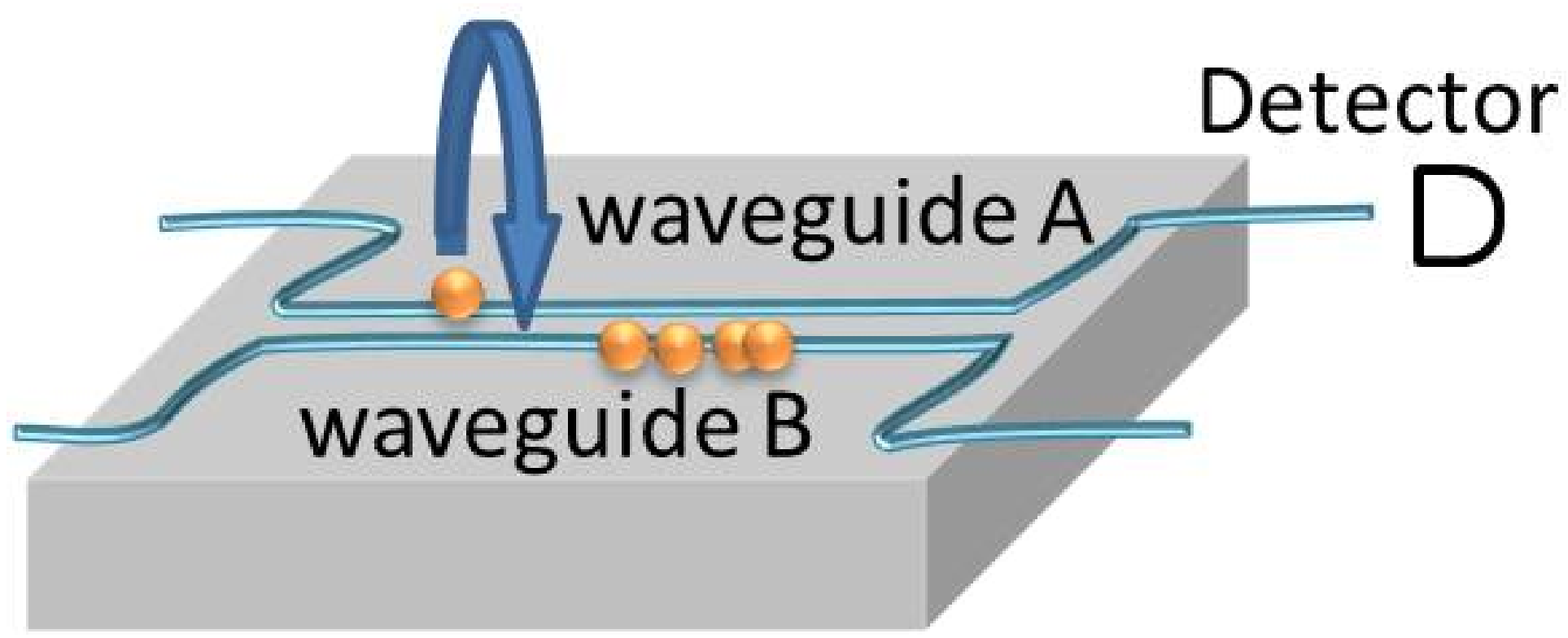} \includegraphics[width=0.18%
\textwidth]{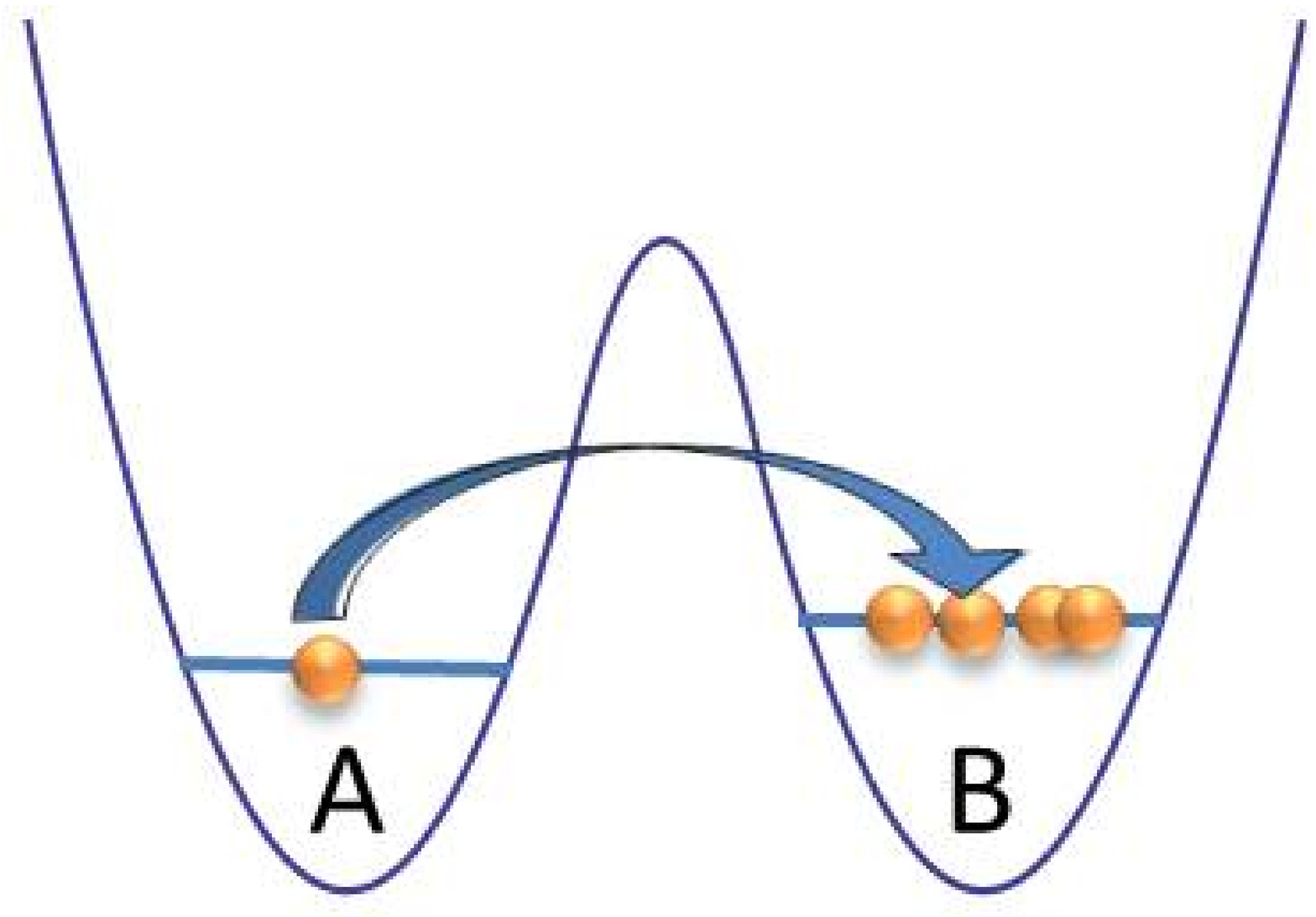}
\caption{(Color online) (a) Silica-on-silicon coupled waveguides (see, for
example, \protect\cite{wg1}). The detector in the dark state at time $t$
signals the tunneling from $A$ to $B$. (b) Tunneling from potential well $A$
to potential well $B$, with the presence of bosons at well $B$. }
\label{Fig0}
\end{figure}

Without loss of generality, we discuss the tunneling of photons between two
coupled single modes in the two
%\textcolor{blue}{by the two \textcolor{blue}{wells, i.e. the tunneling between two potential wells with each contains only one state (Fig. \ref{Fig0}(a)). Experimentally, the tunneling can be realized by two coupled single-mode waveguides labeled $A$ and $B$ (Fig. \ref{Fig0}(b)), for example the silica-on-silicon setting\cite{wg1}.} [\sout{
waveguides labeled $A$ and $B$ [Fig. \ref{Fig0}(a)], which can be realized
using silica on silicon \cite{wg1,wg2}.
%}\textcolor{green}{Ref.A.1,Ref.B.5}]
Arrays of waveguides have been extensively studied with both classical and
quantum light \cite{wg1,wg2}. We can also use fiber couplers. We start with
the simplest possibility that the waveguide $A$ contains one photon and the
waveguide $B$ contains $n$ photons; we calculate the probability $%
P_{(1,n)\rightarrow (0,n+1)}$ of the one photon from the waveguide $A$
tunneling to the waveguide $B$. If $P_{(1,n)\rightarrow
(0,n+1)}>P_{(1,0)\rightarrow (0,1)}$, then we conclude that the presence of $%
n$ photons in waveguide $B$ enhances tunneling. Experimentally, use of an
avalanche photodiode to detect the probability of having zero photons in
waveguide $A$ gives us a measurement of the tunneling of a single photon to
the waveguide $B$. Note that efficient sources of single heralding photons
are available \cite{sp1}. Theoretically, the Hamiltonian for a system of two
coupled waveguides is given by
\begin{equation}
\hat{H}=\hbar \Delta \left( a^{\dagger }a-b^{\dagger }b\right) +\hbar
J\left( a^{\dagger }b+b^{\dagger }a\right) ;  \label{e1}
\end{equation}%
Here $a$, $a^{\dagger }$ ($b,b^{\dagger }$) are the annihilation and
creation operators for the field in $A$ ($B$). Generally, the two waveguides
$A$ and $B$ have different refractive indices leading to different detunings
$\varepsilon _{0}\pm \hbar \Delta $. Here $\varepsilon _{0}$ gives an
overall phase $e^{-\mathtt{i}\varepsilon _{0}t}$, which does not affect the
transport probability. Thus, we set $\varepsilon _{0}=0$. The refractive
index leads to a phase shift of the field but does not change the frequency
of the photon. We consider one-dimensional propagation along the length of
the wave guide. The coupling constant $J$ is the tunneling energy of the
photon needed to go from one waveguide to the other. It depends on the
distance between the two waveguides, and can be calculated from the overlap
of the electric field distributions \cite{bk4}. Its value, say, for two
silicon waveguides at a distance aprat of $4$ $\mu $m is about $0.51$ mm$%
^{-1}$ in units of $\frac{c}{n}$ \cite{wg1}. Other waveguides like AlGaAs
have parameters in a similar range \cite{wg2}.%
%\textcolor{blue}{We'd like to strengthen that here we consider the simple tunneling problems. In the hamiltonian, we do not include any term can change the frequency of photons. The effects due to absorbing/releasing photons are not considered.}[\textcolor{green}{Ref.B.6}]

The Hamiltonian (\ref{e1}) can be diagonalized by writing it in terms of the
angular momentum operators and then by using the properties of the rotation
group. However, in considerations of the dynamics we have found an
alternative and elegant procedure by using the Heisenberg equations of
motion and hence we do not give the exact eigenfunctions of (\ref{e1}). If
the system is initially in the state $\left\vert \Psi \left( 0\right)
\right\rangle =\left\vert n,m\right\rangle $, i.e., there are $n$ photons in
the waveguide $A$ and $m$ photons in the waveguide $B$, then at time $t$
(which is proportional to the propagation distance) the system is in the
state
\begin{equation}
\left\vert \Psi \left( t\right) \right\rangle =\frac{\left( a^{\dagger
}\left( -t\right) \right) ^{n}\left( b^{\dagger }\left( -t\right) \right)
^{m}}{\sqrt{n!m!}}\left\vert 0,0\right\rangle .  \label{e2}
\end{equation}%
Equations (\ref{e2}) can be proved as follows. We write the intial state as%
\begin{equation}
\left\vert \Psi \left( 0\right) \right\rangle =\left\vert n,m\right\rangle =%
\frac{\left( a^{\dagger }\right) ^{n}\left( b^{\dagger }\right) ^{m}}{\sqrt{%
n!m!}}\left\vert 0,0\right\rangle .  \label{a1}
\end{equation}%
Let $U\left( t\right) $ be the time evolution operator, then
\begin{eqnarray}
&&\left\vert \Psi \left( t\right) \right\rangle =U\left( t\right) \left\vert
n,m\right\rangle   \notag \\
&=&\frac{U\left( t\right) \left( a^{\dagger }\right) ^{n}U^{\dagger }\left(
t\right) U\left( t\right) \left( b^{\dagger }\right) ^{m}U^{\dagger }\left(
t\right) }{\sqrt{n!m!}}U\left( t\right) \left\vert 0,0\right\rangle ,
\label{a2}
\end{eqnarray}%
which, on noticing that $\left\vert 0,0\right\rangle $ is an eigenstate of $H
$ with zero energy $U\left( t\right) \left\vert 0,0\right\rangle =\left\vert
0,0\right\rangle $, reduces to (\ref{e2}) with $a^{\dagger }(-t)=U\left(
t\right) a^{\dagger }U^{\dagger }\left( t\right) $, etc. The time dependent
operators $a^{\dagger }(-t)$ and $b^{\dagger }(-t)$ can be obtained through
the Heisenberg equations of motion. We quote the result:
\begin{equation}
\left[
\begin{array}{c}
a^{\dagger }\left( -t\right)  \\
b^{\dagger }\left( -t\right)
\end{array}%
\right] =\left[
\begin{array}{cc}
\sqrt{1-P}e^{-\text{i}\theta } & -\text{i}\sqrt{P} \\
-\text{i}\sqrt{P} & \sqrt{1-P}e^{\text{i}\theta }%
\end{array}%
\right] \left[
\begin{array}{c}
a^{\dagger }\left( 0\right)  \\
b^{\dagger }\left( 0\right)
\end{array}%
\right] .  \label{e3}
\end{equation}%
On defining $\gamma =\frac{\Delta }{J}$, $Q=\sqrt{1+\gamma ^{2}}$, and $%
P_{0}=\frac{1}{Q^{2}}$, the amplitude of the off-diagonal term is given by $%
\sqrt{P}$ with
\begin{equation}
P=P_{0}\sin ^{2}\left( QJt\right) .  \label{e4}
\end{equation}%
The amplitude of diagonal term is given by $\sqrt{1-P}$, and the
corresponding phase $\theta $ can be calculated from the probability $\sqrt{%
1-P}e^{-\text{i}\theta }=\cos \left( QJt\right) -\text{i}\frac{\gamma }{Q}%
\sin \left( QJt\right) $.

At time $t$, the probability that the system is in the state $\left\vert
n^{\prime },m^{\prime }\right\rangle $ is given by $P_{(n,m)\rightarrow
(n^{\prime },m^{\prime })}=\left\vert \left\langle n^{\prime },m^{\prime
}|\Psi \left( t\right) \right\rangle \right\vert ^{2}$. We can then use $%
P_{(n,m)\rightarrow (n^{\prime },m^{\prime })}$ as the time dependent
tunneling probability of $\left( n-n^{\prime }\right) $ photons\ from
waveguide $A$ to waveguide $B$. The tunneling that we discuss is reversible
and is like Josephson tunneling \cite{bk5}. In fact formula (\ref{e4}) is
similar to the relation in the context of Josephson tunneling. We would like
to emphasize that the tunneling discussed here is between two subsystems
that are finite in the tunneling direction.

\section{One-photon tunneling}

\begin{figure}[t]
\includegraphics[width=0.49\textwidth]{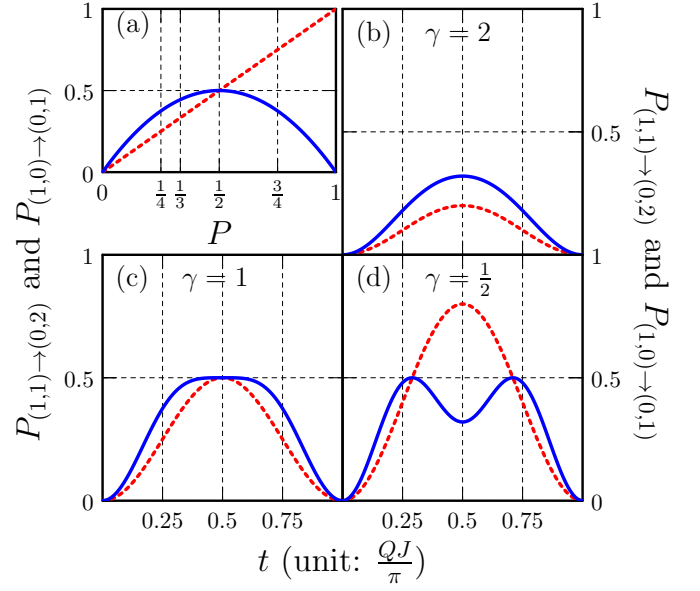}
\caption{(Color online) The tunneling probabilities $P_{(1,0)\rightarrow
(0,1)}$ (red dashed) and $P_{(1,1)\rightarrow (0,2)}$ (blue solid) as
functions of (a) $P$; (b)--(d) scaled time $\frac{QJt}{\protect\pi }$ for
different values of $\protect\gamma =\frac{\Delta }{J}$. }
\label{Fig1}
\end{figure}

If the system is initially in the state $\Psi \left( 0\right) =\left\vert
1,n\right\rangle $, i.e., one photon in waveguide $A$ and $n$ photons in
waveguide $B$, then the tunneling probability of the one photon from $A$ to $%
B$ is given by
\begin{equation}
P_{(1,n)\rightarrow (0,n+1)}=\left( n+1\right) \left( 1-P\right) ^{n}P.
\label{e5}
\end{equation}

The standard case of tunneling is a special case of Eq.(\ref{e5}) when there
is no photon in waveguide $B$, i.e., $n=0$; then $P_{(1,0)\rightarrow
(0,1)}\equiv P$. We note that the quantity $P$ is like the well-known Rabi
oscillation in the populations of a two-level system in an external field.
This is because if there is only one photon in the problem, then the state
space involves only two levels $\left\vert 1,0\right\rangle $ and $%
\left\vert 0,1\right\rangle $. Notice that this one-particle tunneling
probability $P$ is the same as in the fermion case. In fact, we can see
that, when $\Delta $ is large compared to $J$, then the energy gap between
the two modes is large (which means a high barrier at the junction), leading
to a large $\gamma $ and a small tunneling probability. It should be
mentioned that back tunneling $P_{(1,n)\rightarrow (1+m,n-m)}$ with $%
0<m\leqslant n$ exists, but we do not need it as we set the experiment to
detect no photon in $A$.

In order to illustrate how the presence of photons within the other
waveguide affects the tunneling process, we first compare the tunneling
probability $P_{(1,1)\rightarrow (0,2)}$ in the presence of one photon in
the other waveguide with $P$. From Fig. \ref{Fig1}(a), we can see that when $%
P\leq 1/2$, $P_{(1,1)\rightarrow (0,2)}>P$, i.e., the presence of one photon
in the waveguide $B$ enhances the probability of tunneling, and PAT occurs.
It is easy to show from Eq.(\ref{e5}) that the maximum difference $%
(P_{(1,1)\rightarrow (0,2)}-P)_{\mathrm{max}}=1/9$ is reached when $P=1/3$.
%\textcolor{blue}{From Eq.(\ref{e4}), PAT may occurs for any $\Delta$ and $J$, as the PAT condition $P\leq 1/2$ can be always satisfied in some range of time $t$.}
We now show that PAT occurs for any range of values of $\Delta $ and $J$.
When the gap $2\Delta $ is such that $\gamma =\frac{\Delta }{J}\geq 1$, we
have $P_{0}<1/2$, $P<1/2$, and PAT always occurs. This is shown in Fig. \ref%
{Fig1}.(b). Here $\gamma =2$ and $P_{(1,1)\rightarrow (0,2)}\geq P$ for any
time $t$. When the energy gap is small $\gamma \leq 1$ and thus $P_{0}\geq
1/2$, we can still observe PAT. The oscillation structure of $P$, Eq. (\ref%
{e4}) guarantees that there exist time regions such that $P\leq 1/2$ so that
$P_{(1,1)\rightarrow (0,2)}\geq P$. [Figs. \ref{Fig1}(c) and \ref{Fig1}(d)].

From Fig. \ref{Fig1}(a), we also observe the appearance of photon-inhibited
tunneling when $P>1/2$. In particular, when $P=1$ ($100\%$ tunneling
probability without the presence of any photon in waveguide $B$), we have
exactly $P_{(1,1)\rightarrow (0,2)}=0$, and the photon tunneling is totally
inhibited. PIT can be observed only when the gap is small, $\gamma <1$, so
that $P_{0}>1/2$, which allows $P>1/2$ in some time region [Fig. \ref{Fig1}%
(d)].

\begin{figure}[t]
\includegraphics[width=0.49\textwidth]{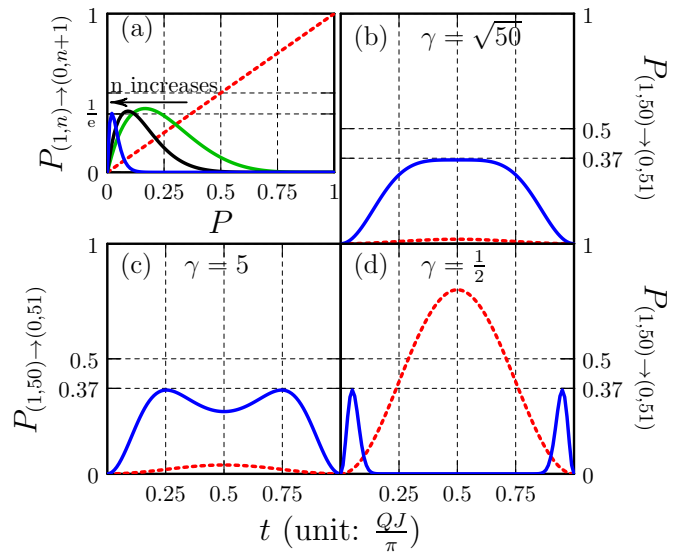}
\caption{(Color online) (a) The tunneling probability $P_{(1,n)\rightarrow
(0,n+1)}$ as a function of $P$ for $n=0$ (red dashed); $n=5$ (green); $n=10$
(black); $n=50$ (blue). (b)--(d) The time dependence of the tunneling
probability $P_{(1,50)\rightarrow (0,51)}$ (blue) as compared to that of $%
P_{(1,0)\rightarrow (0,1)}$ (red dashed) for different values of $\protect%
\gamma $. }
\label{Fig2}
\end{figure}

The existence of PAT and PIT shows a competition mechanism introduced by the
bosonic nature of photons, as is seen from Eq. (\ref{e5}). Two factors are
multiplied with $P$: the first one $\left( n+1\right) \geq 1$ is an aided
term, which makes it possible for $P_{(1,n)\rightarrow (0,n+1)}$ to be
bigger than $P$. It comes from the bosonic nature that $a^{\dag }\left\vert
n\right\rangle =\sqrt{n+1}\left\vert n+1\right\rangle $. The bosonic nature
shows that if a state contains more bosons, then it is easier to add an
extra boson to it. The second term $\left( 1-P\right) ^{n}\leq 1$ is an
inhibited term, which reflects the tendency that the $n$ photons from
waveguide $B$ stay in waveguide $B$. Notice that $\left( 1-P\right) $ gives
the probability that the one photon still wants to stay at its own site. The
power $n$ reflects the probability of the $n$-photon state remaining an $n$%
-photon state. When $n$ is larger, this term becomes smaller. The aided term
$\left( n+1\right) $ is fixed when the photon number is fixed; while the
inhibited term $\left( 1-P\right) ^{n}$ can decrease from $1$ to $0$ when $P$
increases from $0$ to $1$. As a result: if $P$ is small, the aided term
dominates and we observe PAT; if $P$ is large, then we observe PIT.

We next examine $P_{(1,1)\rightarrow (0,2)}$. When $P<1/2$, the tunneling
probability is small; every photon wants to stay in its own state, and the
inhibited term $\left( 1-P\right) ^{n}$ is large. At this time the aided
term $(n+1)$ is important, leading to $P_{(1,1)\rightarrow (0,2)}>P$. When $%
P>1/2$, the inhibited term suppresses the positive effect of the aided term $%
(n+1)$.

We now discuss the general case of the tunneling probability $%
P_{(1,n)\rightarrow (0,n+1)}$ of one photon tunneling in the presence of $n$
photons. From Fig. \ref{Fig2}(a), we can see that when the number of photons
is increased, the region of PAT becomes smaller and occurs at smaller values
of $P$; however, the maximum value does not decrease very much. In fact,
from Eq.(\ref{e5}), we get $(P_{(1,n)\rightarrow (0,n+1)})_{\mathrm{max}}=(1-%
\frac{1}{n+1})^{n}$ when $P=1/(n+1)$. When the number of photons in
waveguide $B$ is very large, $n\rightarrow \infty $, we can get $%
(P_{(1,n)\rightarrow (0,n+1)})_{\mathrm{max}}\rightarrow 1/\mathrm{e}\simeq
0.37$. Therefore we get very significant PAT especially when $\Delta $ is
large, which leads to very small values of $P$. We compare the tunneling
probability $P_{(1,n)\rightarrow (0,n+1)}$ with $P$ for different $\gamma $.
In Fig. \ref{Fig2}(b), when $\gamma =\sqrt{n}=\sqrt{50}$ so that $P_{0}=1/51$%
, we can see that the tunneling without any photons in waveguide $B$ is very
small, while $P_{(1,50)\rightarrow (0,51)}$ has a plateau at $0.37$ for a
large time range; In Fig. \ref{Fig2}(c), the aided tunneling is still
significant for $\gamma =5$. In Fig. \ref{Fig2}(d), when the gap is small,
the aided tunneling acts as a pulse in the vicinity of time period $T$. This
means that even for a large gap $\Delta $, we can always find a finite
photon tunneling probability $(P_{(1,n)\rightarrow (0,n+1)})_{\mathrm{max}}$
near $\frac{1}{\mathrm{e}}$ by choosing $n\geq \gamma ^{2}=\frac{\Delta ^{2}%
}{J^{2}}$, whereas for large $\Delta $, $P_{(1,0)\rightarrow (0,1)}$ is
negligible.

\section{Multi-photon tunneling}

\begin{figure}[t]
\includegraphics[width=0.4\textwidth]{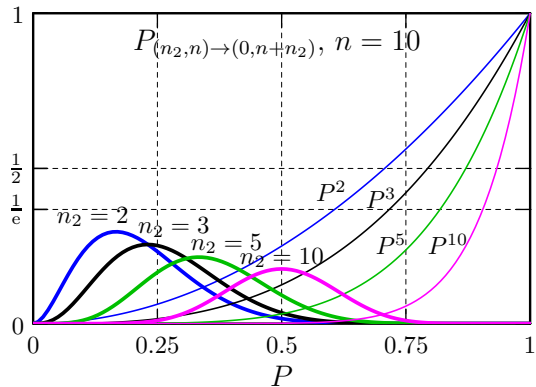}
\caption{$P_{(n_{2},n)\rightarrow (0,n+n_{2})}$ (thick lines), the
multiphoton tunneling probability in presence of $10$ photons in the
waveguide $B$, as a function of $P$ for $n_{2}=2,3,5,10$. The multi-photon
tunneling probabilities without presence of photons in the waveguide $B$, $%
P_{(n_{2},0)\rightarrow (0,n_{2})}=P^{n_{2}}$ are given by thin lines. }
\label{Fig3}
\end{figure}

The above PAT can be generalized to multi-photon tunneling. In Fig. \ref%
{Fig3}, we compare $P_{(n_{2},n)\rightarrow (0,n+n_{2})}$ with $%
P_{(n_{2},0)\rightarrow (0,n_{2})}=P^{n_{2}}$. Obviously, $%
P_{(n_{2},0)\rightarrow (0,n_{2})}$ decreases when $n_{2}$ increases since $%
P<1$. This is intuitive in that it is harder for more photons to tunnel to
another waveguide. We can see that when $n_{2}=10$, the probability $%
P_{(n_{2},0)\rightarrow (0,n_{2})}$ is close to zero for most values of $P$.
However, with photons in waveguide $B$, the tunneling is significant. We can
always find a finite maximum of $P_{(n_{2},n)\rightarrow (0,n+n_{2})}$. If $%
n\gg n_{2}$, this tunneling probability is even greater than the one-photon
tunneling $P_{(1,0)\rightarrow (0,1)}$ as shown in Fig. \ref{Fig3}.

Using Eq. (\ref{e2}), we find the result for $P_{(n_{2},n)\rightarrow
(0,n+n_{2})}$ to be:
\begin{equation}
P_{(n_{2},n)\rightarrow (0,n+n_{2})}=\left(
\begin{array}{c}
n+n_{2} \\
n_{2}%
\end{array}%
\right) \left( 1-P\right) ^{n}P^{n_{2}}.  \label{e6}
\end{equation}%
where $\left(
\begin{array}{c}
n \\
m%
\end{array}%
\right) =\frac{n!}{(n-m)!m!}$ is the binomial coefficient. From this
equation, we can clearly see that the inhibited term is still given by $%
\left( 1-P\right) ^{n}$, while the aided term is changed to $\left(
\begin{array}{c}
n+n_{2} \\
n_{2}%
\end{array}%
\right) $. This means that aided tunneling should be more important when $%
n_{2}$ increases. However, noticing that $P_{(n_{2},0)\rightarrow (0,n_{2})}$
decreases, the absolute value $P_{(n_{2},n)\rightarrow (0,n+n_{2})}$ may
still decrease. This can be seen from Fig. \ref{Fig3} which shows the
decreasing peak of $P_{(n_{2},n)\rightarrow (0,n+n_{2})}$ with the increase
of $n_{2}$. The maximum peak value is given by
\begin{equation}
\left( P_{(n_{2},n)\rightarrow (0,n+n_{2})}\right) _{\max }=\left(
\begin{array}{c}
n+n_{2} \\
n_{2}%
\end{array}%
\right) \frac{n_{2}^{n_{2}}n^{n}}{\left( n+n_{2}\right) ^{n+n_{2}}}.
\label{e7}
\end{equation}%
which is reached when $P=\frac{n_{2}}{n+n_{2}}$. The limit when $%
n\rightarrow \infty $, is the maximum of $\frac{1}{n_{2}!}(\frac{n_{2}}{%
\mathrm{e}})^{n_{2}}$. It decreases when $n_{2}$ increases. However, even
for $n_{2}=10$, we still have the finite tunneling at about $12.5\%$, which
is much greater than the corresponding tunneling probability $%
P_{(10,0)\rightarrow (0,10)}=\left( \frac{1}{20}\right) ^{10}$ in the
absence of any photons in the waveguide $B$.

\section{The field in a coherent state and a squeezed state}

In an experiment, it is much easier to prepare the field in a coherent state
than in a state with fixed photon number. The coherent state $\left\vert
\beta \right\rangle =e^{-\left\vert \beta \right\vert
^{2}/2}\sum_{n=0}^{\infty }\frac{\beta ^{n}}{\sqrt{n!}}\left\vert
n\right\rangle $ has the average photon number $\bar{n}=\left\langle \beta
\right\vert b^{\dag }b\left\vert \beta \right\rangle =|\beta |^{2}$. We may
then discuss the possibility of PAT and PIT with the field in a coherent
state in the waveguide $B$ with a fixed average photon number $\bar{n}$. The
probability $P_{\left( n_{2};\beta \right) \rightarrow \left( 0;\beta
,n_{2}\right) }$ of $n_{2}$ photons tunneling from waveguide $A$ to
waveguide $B$ is given by (see the Appendix)
\begin{equation}
P_{\left( n_{2};\beta \right) \rightarrow \left( 0;\beta ,n_{2}\right) }=e^{-%
\bar{n}}\left. _{1}F_{1}\right. \left[ 1+n_{2},1,\bar{n}\left( 1-P\right) %
\right] P^{n_{2}},  \label{e8}
\end{equation}%
where the confluent hypergeometric function $\left. _{1}F_{1}\right. \left[
a,b;z\right] $ is defined by%
\begin{equation}
\left. _{1}F_{1}\right. \left[ a,b;z\right] =\sum_{k-0}^{\infty }\frac{%
\left( a\right) _{k}}{\left( b\right) _{k}}\frac{z^{k}}{k!},
\end{equation}%
here $\left( a\right) _{k}$ and $\left( b\right) _{k}$ are Pochhammer
symbols, given by $\left( a\right) _{0}=1$ and $\left( a\right) _{k}=a\left(
a+1\right) \cdots \left( a+k-1\right) $ for $k>1$.

\begin{figure}[tbp]
\includegraphics[width=0.49\textwidth]{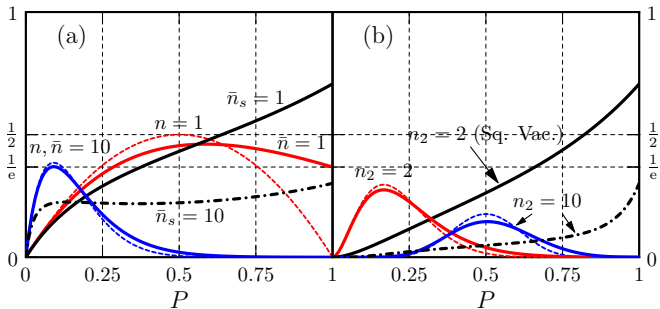}
\caption{(Color online) Comparison of photon tunneling probabilities with
the field in waveguide $B$ in a coherent state and in a Fock state. (a) The
single-photon tunneling probability $P_{(1;\protect\beta )\rightarrow (0;%
\protect\beta ,1)}$ (solid lines) and $P_{(1;n)\rightarrow (0;n+1)}$ (dashed
lines) as a function of $P$; (b) The multiphoton tunneling probability $%
P_{(n_{2};\protect\beta )\rightarrow (0;\protect\beta ,n_{2})}$ (solid
lines) and $P_{(n_{2},n)\rightarrow (0,n+n_{2})}$ (dashed lines) in the
presence of (on average) $ten$ photons in waveguide $B$, as a function of $P$%
. The black solid and dash-dotted lines represent the tunneling probability
for a field in squeezed vacuum in waveguide $B$.}
\label{Fig4}
\end{figure}

In the special case of one-photon tunneling, the result is rather simple:
\begin{equation}
P_{\left( 1;\beta \right) \rightarrow \left( 0;\beta ,1\right) }=e^{-\bar{n}%
P}\left[ 1+\bar{n}\left( 1-P\right) \right] P.  \label{e9}
\end{equation}%
The formulas (\ref{e8}) and (\ref{e9}) are complicated; it is not easy to
separate aided and inhibited terms. However, from Fig. \ref{Fig4}, we
observe that, the results for the coherent state case are very similar to
those for the case of a Fock state in waveguide $B$, especially when $\bar{n}%
=n$ is large. We examine the maximum of $P_{\left( 1;\beta \right)
\rightarrow \left( 0;\beta ,1\right) }$, which is $\frac{\sqrt{\bar{n}^{2}+2%
\bar{n}+5}-2}{\bar{n}}\mathrm{e}^{\frac{1}{2}\left( \sqrt{\bar{n}^{2}+2\bar{n%
}+5}-\bar{n}-3\right) }$, and occurs for $P=\frac{\bar{n}^{2}+3\bar{n}-\bar{n%
}\sqrt{\bar{n}^{2}+2\bar{n}+5}}{2\bar{n}^{2}}$. For large $\bar{n}$, $\lim_{%
\bar{n}\rightarrow \infty }\left( P_{\left( 1;\beta \right) \rightarrow
\left( 0;\beta ,1\right) }\right) _{\mathrm{max}}=\frac{1}{\mathrm{e}}$,
which is the same as for the Fock state.

The tunneling probabilities are also sensitive to the photon statistics of
photons in waveguide $B$. To illustrate this, we consider the field in a
squeezed state $\left\vert \xi \right\rangle =\frac{1}{\sqrt{\cosh r}}%
\sum_{n=0}^{\infty }e^{\text{i}n\varphi }\left( \tanh r\right) ^{n}\frac{%
\sqrt{\left( 2n\right) !}}{n!2^{n}}\left\vert 2n\right\rangle $ with $%
\overline{n}_{s}=\sinh ^{2}r=10$. In this case, calculations show that (see
the Appendix)
\begin{equation}
P_{\left( n_{2};\xi \right) \rightarrow \left( 0;\xi ,n_{2}\right) }=\frac{%
P^{n_{2}}\left. _{2}F_{1}\right. [\frac{1+n_{2}}{2},\frac{2+n_{2}}{2},1;%
\frac{\left( 1-P\right) ^{2}\overline{n}}{1+\overline{n}}]}{\sqrt{1+%
\overline{n}}},  \label{e10}
\end{equation}%
where $\left. _{2}F_{1}\right. [a,b,c;z]$ is the hypergeometric function,
defined by Pochhammer symbols:
\begin{equation}
\left. _{2}F_{1}\right. [a,b,c;z]=\sum_{k-0}^{\infty }\frac{\left( a\right)
_{k}\left( b\right) _{k}}{\left( c\right) _{k}}\frac{z^{k}}{k!}.
\end{equation}%
The dash-dotted black line in Fig. \ref{Fig4} shows the behavior of (\ref%
{e10}) for $n_{2}=1$ and $10$. The behavior is clearly different from the
case of a coherent state: a long plateau occurs. We can conclude that the
tunneling probability in the presence of a field in squeezed state $%
P_{\left( n_{2};\xi \right) \rightarrow \left( 0;\xi ,n_{2}\right) }$ is
mostly inhibited compared to the case when no field is present in waveguide $%
B$.

\section{Conclusion}

In conclusion we have shown how the tunneling of a single photon as well as
multiphoton tunneling can be enhanced or inhibited by the presence of
photons. We presented the physical reasons behind such an enhancement or
inhibition. A crucial role is played by the bosonic nature of photons. The
waveguide structures or fiber couplers are known to be almost
decoherence-free; however, if need be then the decoherence effects can be
taken into account\cite{con}. Although we explicitly considered the simplest
case of a coupler the results can be extended to arrays of couplers.
Tunneling is a universal effect in physics; and therefore, the results of
this paper would be applicable to all situations in which bosons are
involved. Further, the results of this paper should have a bearing on the
quantum walk of a single photon in the presence of other photons \cite{QW}.
The results of this paper should also be applicable to other bosonic systems
like cold atoms in multiple traps.

X. L. would like to acknowledge Amanda Taylor for a careful reading of the
manuscript.

%\appendix

\section*{Appendix: Derivation of Photon Tunneling for Field in Coherent and Squeezed
States}

Consider first the case when the field in the waveguide is in a coherent
state $\left\vert \beta \right\rangle =\sum_{m=0}^{\infty }c_{m}\left\vert
m\right\rangle $ with $c_{m}=e^{-\left\vert \beta \right\vert ^{2}/2}\frac{%
\beta ^{m}}{\sqrt{m!}}$. Thus the initial state is $\left\vert n_{2};\beta
\right\rangle $ instead of $\left\vert n,m\right\rangle $ as considered in
Sec. \ref{s2}. The wave function at time $t$ is%
\begin{equation}
\left\vert \Psi \left( t\right) \right\rangle =U\left( t\right) \left\vert
n_{2};\beta \right\rangle =\sum_{m=0}^{\infty }c_{m}U\left( t\right)
\left\vert n_{2},m\right\rangle ,  \label{ae1}
\end{equation}%
where $U\left( t\right) \left\vert n_{2},m\right\rangle $ can be obtained
from Eqs. (\ref{a2}) - (\ref{e4}). The probability for $n_{2}$ photon
tunneling is given by%
\begin{equation}
P_{\left( n_{2};\beta \right) \rightarrow \left( 0;\beta ,n_{2}\right)
}=\sum_{l=0}^{\infty }\left\vert \left\langle 0,l\right. \left\vert \Psi
\left( t\right) \right\rangle \right\vert ^{2}=\sum_{l=0}^{\infty
}\left\vert c_{l}\right\vert ^{2}P_{\left( n_{2},l\right) \rightarrow \left(
0,l+n_{2}\right) },  \label{ae2}
\end{equation}%
where $P_{\left( n_{2},n\right) \rightarrow \left( 0,n+n_{2}\right) }$ is
given by Eq. (\ref{e6}) and hence
\begin{equation}
P_{\left( n_{2};\beta \right) \rightarrow \left( 0;\beta ,n_{2}\right)
}=\sum_{l=0}^{\infty }\left\vert c_{l}\right\vert ^{2}\left(
\begin{array}{c}
l+n_{2} \\
n_{2}%
\end{array}%
\right) \left( 1-P\right) ^{l}P^{n_{2}},  \label{ae3}
\end{equation}%
The series in Eq. (\ref{ae3}) can be summed up, leading to the result (\ref%
{e8}).

If the field in the waveguide is in the squeezed state $\left\vert \xi
\right\rangle =\frac{1}{\sqrt{\cosh r}}\sum_{m=0}^{\infty }e^{\text{i}%
m\varphi }\left( \tanh r\right) ^{m}\frac{\sqrt{\left( 2m\right) !}}{m!2^{m}}%
\left\vert 2m\right\rangle $, then Eq. (\ref{ae3}) holds with $c_{2l}=\frac{1%
}{\sqrt{\cosh r}}e^{\text{i}n\varphi }\left( \tanh r\right) ^{n}\frac{\sqrt{%
\left( 2n\right) !}}{n!2^{n}}$ and $c_{2l+1}=0$. On substituting these
values and summing the series we get the result (\ref{e10}). %

\end{document}